\documentclass{elsart}
\usepackage {graphicx,amsmath}
\begin{document}

\begin{center}
 {\Large\bf  Quantum theory without logical paradoxes?  } \\[2mm]
            Milo\v{s} V. Lokaj\'{\i}\v{c}ek    
\footnote{e-mail: lokaj@fzu.cz} \\
Institute of Physics of the AS CR, v.v.i., 18221 Prague 8, Czech Republic  \\

\end{center}
\vspace{3mm}

{\bf Abstract } \\
In contribution published in AIP Conference Proceedings, No. 1018 (p. 40-5) some discrepancies in Copenhagen quantum mechanics have been summarized and arguments have been gathered why in the description of microscopic reality hidden-variable theory should be preferred. It has been shown, too, that the earlier gap in physical descriptions of microscopic and macroscopic worlds has been practically removed. 
All corresponding statements have been based on proofs that have been published earlier and reproduced in the given contribution only partially. In the present paper the whole approach will be explained more systematically. Some important points newly analyzed will be added.   \\ 
 

\section{Introduction}
\label{sec1}
In preceding papers of ours (see mainly \cite{conc1,adv}) the arguments have been proved leading to the conclusion that the hidden-variable (HV) theory should be preferred to the Copenhagen quantum mechanics. The results based on these arguments have been presented to the conference held in Trieste (January 2008) and published in corresponding Proceedings \cite{ffp9} (similar results being contained also in Ref. \cite{hradec}). In the present paper the more systematic explanation of the whole approach will be provided. 

The new results may be shortly summarized: Two different (Copenhagen and ensemble) interpretations of quantum-mechanical model discussed earlier have represented in fact two divers theories differing in important assumptions. The ensemble alternative being identical practically with mere Schr\"{o}dinger equation (and equivalent to hidden-variable theory) refuses in principle physically interpreted superposition principle and requires to extend corresponding Hilbert space to be in full harmony with time-dependent solutions of the equation proposed by Schr\"{o}dinger \cite{schr}. All previous  criticisms (Pauli, Susskind and Glogover, Einstein) often discussed have been removed in such a case. And no logical contradictions and paradoxes are involved more in the description of matter reality. It has been also proved that the basic solutions (represented always by one Hamilonian eigenfunction) of Schr\"{o}dinger equation lead to the same results as classical physics. However, some classically possible states do not seem to correspond to basic Schr\"{o}dinger solutions, which concerns the existence of discrete states. It is also possible to say that the microscopic physics may provide now the same ontological picture as it is known from classical physics.   

Main assumptions the Copenhagen quantum mechanics has been based on will be introduced in Sec. \ref{sec2} and the difference between two earlier often discussed interpretation alternatives will be explained. Critical comments of Pauli \cite{pauli} and others discussed during the second half of the 20th century will be explained in Sec. \ref{sec3}. The necessity of extended Hilbert space and time irreversibility will be then handled in Sec. \ref{sec4}. The story of EPR experiment will be described and explained in Sec. \ref{sec5}. And in Sec. \ref{sec6} the physical meanings of different limits of Bell's operator will be shown. The possibility of applying the HV (hidden-variable) theory to whole physical reality will be discussed in Sec. \ref{sec7}. The results and consequences of the experiment with three polarizers will be presented in Sec. \ref{sec8}.     
The actual relation between the Schr\"{o}dinger equation and classical physics will be then shown in Sec. \ref{sec9}. The structure of Hilbert space corresponding to the HV theory will be discussed in Sec. \ref{sec10}.
Some concluding remarks will be introduced in Sec. \ref{sec11}.

\section { Copenhagen quantum mechanics and basic assumptions }
\label{sec2}
To understand the whole problem it is necessary to start with main assumptions on which the Copenhagen quantum mechanics is based. Leaving aside some technical assumptions it is possible to introduce four following ones:  

 - first of all it is the validity of time-dependent Schr\"{o}dinger equation \cite{schr} 
 \begin{equation}
  i\hbar\frac{\partial}{\partial t}\psi(x,t)=H\psi(x,t), \;\;\;\;
     H=-\frac{\hbar^2}{2m}\triangle + V(x)   \label{schr}
\end{equation}
where Hamiltonian $H$ represents the total (kinetic and potential) energy of a given physical system and $x$ represents the coordinates of all matter objects; 

 - the evolution of a physical system is described by function $\psi(x,t)$ and all physical quantities may be expressed as the expected values of corresponding operators:
 \begin{equation}
           A(t) = \int\psi^*(x,t)A_{op}\psi(x,t)dx    
\end{equation}
where $A_{op}$ and functions $\psi(x,t)$ may be represented by operators and vectors in a suitable Hilbert space;

 - in the case of Copenhagen quantum mechanics it has been required for the corresponding Hilbert space to be spanned on one set of Hamiltonian eigenfunctions $\psi_E(x)$:
\begin{equation}
                H\psi_E(x) = E\psi_E(x);
\end{equation} 

 - and in addition to, the mathematical superposition principle valid in any Hilbert space has been interpreted in physical sense, i.e., any superposition of two vectors has represented again another pure (basic) physical state. 

It has been spoken often about two different interpretation alternatives of the quantum-mechanical mathematical model: orthodox (or Copenhagen) and statistical (or ensemble). However, it has been never sufficiently stressed that both these alternatives have not concerned identical model and have corresponded to the different sets of basic assumptions. While the Copenhagen alternative has involved all four preceding assumptions the ensemble alternative (denoted usually also as HV theory) has corresponded to the first two assumptions only (being equivalent in principle to the mere Schr\"{o}dinger equation). And it is necessary to speak about two different theories, differing significantly in their properties as well as in assumptions.

 The Copenhagen alternative has been often denoted as supported by different experimental data. However, in all such cases only the assumption of the mere Schr\"{o}dinger equation (i.e., of the HV theory) has been tested; without the last two assumptions (forming Copenhagen alternative) having been actually involved. All four assumptions have been involved practically only in interpreting the EPR experiment, which will be discussed in Sec. \ref{sec5}. 

 As to the HV theory the Hilbert space must be chosen according to corresponding physical system; it must be always extended (at least doubled) in comparison to the Hilbert space required by the third assumption. In such a suitably extended Hilbert space the critical comments of Pauli and also of Susskind and Glogover are not more valid (see the next section). 

\section { Critical comments of Pauli and others }
\label{sec3}
The Copenhagen quantum mechanics was accepted as valid by physical community even if it exhibited some logical and ontological paradoxes and some critical comments were brought against its full regularity.
Already in 1933 Pauli \cite{pauli} showed that under the validity of all assumptions introduced in Sec. \ref{sec2} it was necessary for the corresponding Hamiltonian to possess continuous energy spectrum from $-\infty$ to $+\infty$, which disagreed with the fact that the energy was defined as positive quantity, or at least limited always from below.
In 1964 Susskind and Glogover \cite{suss} showed then that exponential phase operator was not unitary, which indicated that the given Hilbert space was not fully complete to represent a corresponding physical system quite regularly. Many attempts have been done to solve these deficiencies during the 20th century. The reason of having been unsuccessful may be seen in the fact that practically in all cases both the shortages were regarded and being solved as one common problem.

The corresponding solution seems to have been formulated only recently (see Refs. \cite{kulo1,kulo2}) when it has been shown that it is necessary to remove two mentioned shortages one after the other. As to the simple system of two free colliding particles the criticism of Pauli may be removed if the Hilbert space required by the third assumption has been doubled as proposed by Lax and Phillips in 1967 (see \cite{lax1,lax2}); it consists then at least of two mutually orthogonal subspaces ($\mathcal{H} = \Delta^- \oplus \Delta^+$); each of them being spanned on one basis of Hamiltonian eigenfunctions (see also Sec. \ref{sec4}). 

As to the non-unitarity of exponential phase operator it is necessary to mention that this operator was discussed for the first time by Dirac \cite{dirac} in the case of linear harmonic oscillator (i.e.,  $V(x)=kq^2$). Annihilation and creation operators 
\vspace{-4mm}
\begin{equation}
    a \;=\; p - im\omega q, \;\;\; a^{\dag} \;=\;  p+im\omega q,
                    \;\;\;\;\: \omega=\sqrt{\frac{k}{m}}
\end{equation}
\vspace{-1mm}
were introduced, fulfilling the relations
\begin{equation}
    [H,a] \;=\; -\omega a\,, \;\;\; [H, a^\dag] \;=\; \omega  a^{\dag}\,.
\end{equation}
 It was then possible to define operator
\begin{equation}
 \mathcal{E}\,=\, (a a^\dag+1)^{1/2}a\:,
             \;\;\;  \mathcal{E}^\dag  \;=\; a^\dag (aa^\dag +1)^{1/2})
\end{equation}
fulfilling the relations
\begin{equation}
 [H,\mathcal{E}] \;=\; -\omega \mathcal{E}\;, \;\;\;
          [H,\mathcal{E}^\dag] \;=\; +\omega \mathcal{E}^\dag 
 \end{equation}
and representing exponential phase operator defined as
\begin{equation}
  \mathcal{E} \;=\;e^{-i\omega\Phi}      \label{exph}
\vspace{-2mm}
\end{equation}
where $\Phi$ is the phase.

And it was shown later by Susskind and Glogower \cite{suss}  that in the Hilbert space corresponding to the third assumption the operator  $\mathcal{E}$ was not unitary, but only isometric, as it held
 $\;\mathcal{E}^\dag\mathcal{E}\:u_{1/2} \,\equiv\,  0$. The unitarity condition has not been fulfilled for the state vector corresponding to the minimum-energy (or vacuum) state. 
The Hilbert space (extended to solve Pauli's problem) should be further doubled and formed by combining two mutually orthogonal subspaces corresponding to systems with opposite angular momentums. 
They should be bound together by the added action of exponential phase operator, linking together the corresponding vacuum states as it was proposed already by Fain \cite{fajn}; see also \cite{kulo1,kulo2}. 
In any case, the given non-unitarity of exponential phase operator has represented always a less substantial problem than the criticism of Pauli, as it has concerned the completeness of the Hilbert space and not any actual discrepancy.  

Very important criticism was delivered, however, by Einstein and collaborators \cite{einst} who proposed the so called EPR Gedankenexperiment to argue that some action at macroscopic distance between microscopic particles should exist in the Copenhagen model. The EPR problem has been repeatedly discussed, which has been accompanied by some important mistakes leading to false conclusions The whole story and contemporary solution of the given problem will be described in Sec. \ref{sec5}. 

However, before ending the present section another criticism should  be yet added. It concerns the existence of discrete states in Schr\"{o}dinger equation when the two last assumptions are added. It is evident that in such a case all mathematical superpositions should represent mutually equivalent physical states and practically no quantized (discrete) states should exist in experimental reality. The given problem has been removed in the HV theory as only eigenstates belonging to Hamiltonian eigenvalues represent now "pure" states  and any superposition of theirs represents statistical mixture.

It means that the structure of the Hilbert space in the HV theory differs significantly from that required in the case of the Copenhagen quantum mechanics. It consists always of mutually orthogonal subspaces in the dependence on the type of a corresponding physical system (see Sec. \ref{sec4} and eventually also Sec. \ref{sec10}). The time flow is then described as irreversible, which will be demonstrated on the physical situation corresponding to the evolution of two-particle system being described in the next section. 

\section {  Hidden-variable theory and irreversible time flow      }
\label{sec4}

The HV theory differs from the Copenhagen quantum mechanics also in one further important point; the time evolution is not more reversible, which follows simply from the Schr\"{o}dinger equation and the extended Hilbert space. 

Let us suppose now that the given physical system consists of two particles. It may be described in its center mass system (CMS) with the help of Hamiltonian
\vspace{-2mm}
\begin{equation}
  H \;=\; \frac{p^2}{2m} \;+\; V(x)            \label{ham}
\end{equation}
where $m$ is reduced mass, $p$ - momentum of one particle, $V(x)$ is mutual potential between the given particle pair and $x$ represents the positions (mutual distance) of particles. The Hilbert space $\mathcal H$ corresponding to the HV theory consists then of two subspaces:
\begin{equation}
      \mathcal{H}  \;\equiv\;  \{\Delta^-,\Delta^+\}
\end{equation}
that are mutually related by evolution operator
\begin{equation}
            U(t) \;=\; e^{-iHt} \;\;  (t \ge 0).
\end{equation}
 It holds, e.g., (see \cite{lax1,lax2})
\begin{equation}
\mathcal{H} \;=\; \overline{\Sigma_t U(t)\Delta^-}
                             \;=\;\overline{\Sigma_t U(-t)\Delta^+}.
\end{equation}

Individual subspaces $\Delta^-$ and $\Delta^+$ are spanned on one set of
Hamiltonian eigenfunctions in usual way. However, the superposition principle cannot be applied to in the physical sense. The basic physical states are represented by Hamiltonian eigenstates only; their superpositions represent statistical combinations of basic states. Any $t$-dependent 
function $\psi(x,t)$ obtained by solving Schr\"{o}dinger equation may be then
represented by a trajectory characterized by given initial
conditions.

In case of continuous Hamiltonian spectrum (two free particles) any point on such a trajectory may be characterized by expectation values of the operator   
\vspace{-2mm}
  $$ R\;=\;\frac{1}{2}\{{\bf p}.{\bf q}\}, \;\;\; \langle i[H,R]\rangle >0 $$
where ${\bf q}$ and ${\bf p}$  are coordinates and momentum
components of one particle in CMS. The states belonging to
$\Delta^-$ are incoming states, and those of $\Delta^+$ - outgoing
states (independently of chosen coordinate system). The evolution goes always in one direction from "in"-subspace to "out"-subspace (for more details see \cite{lok98}).

As these two different kinds of states represent quite different experimental situations it is necessary to separate "in" and "out" states into
two mutually orthogonal subspaces. It is then also possible to join an additional orthogonal subspace that might represent corresponding resonance formed in a particle collision (or an unstable particle decaying into the given particle pair), i.e.
\begin{equation}
 \mathcal{H} \;\equiv\; \{\Delta^-\oplus\Theta\oplus\Delta^+\}\; ;
\end{equation}
see also Ref. \cite{alda} where the corresponding extended Hilbert space was derived independently as the consequence of exact exponential decay law.
It is only necessary to define the action of evolution operator
between $\Theta$ and other subspaces in agreement with evolution
defined already in individual $\Delta^\pm$-subspaces.
The evolution goes  in one direction, at least from the global point of
view; evolution processes between internal states of $\Theta$ may be
rather chaotic. However, global trajectories tend always in
one direction;  see the scheme in Fig. 1.

\vspace{5mm}
\begin{figure}[htb]
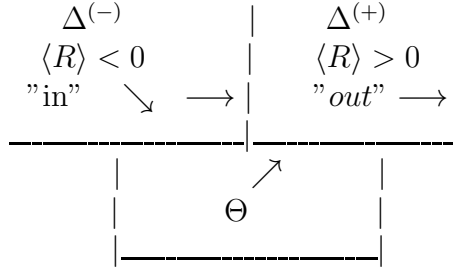

\begin{center}
 \hspace*{-0.1cm}   $\Delta^{(-)}\; \hspace{1.54cm} |
                                            \hspace{0.8cm} \;\Delta^{(+)}$ \\
 \hspace*{-0.1cm}  { $\langle R\rangle <0 \hspace{1.33cm} |
                                      \hspace{0.8cm}\langle R\rangle >0$  \\
 \hspace*{0.1cm} { "in" } \hspace{0.1cm} $\searrow
                                   \hspace{0.4cm}     \longrightarrow  |
                               \hspace{0.8cm} { "out" } \longrightarrow$ \\
 \hspace*{.05cm}  $ \_\_\_\_\_\_\_ \_\_\_\_\_\_\_\_\_\_\_\_\_\_|
            \_\_\_\_\_\_\_\_\_\_\_\_\_\_\_\_\_\_ $ \\
   \hspace*{0.7cm}$| \hspace{1.6cm}  \nearrow \hspace{1.15cm}  | $    \\
 \hspace*{0.64cm}$| \hspace{1.4cm} { \Theta}\hspace{1.7cm}| $      \\
 \hspace*{0.8cm}$|\_\_\_\_\_\_\_\_\_\_\_\_\_\_\_\_\_\_\_\_\_\_\_|    $ }      
\caption { \it { Scheme of the Hilbert space (for a two-particle system)
extended according to original proposal of Lax and Phillips; three mutually orthogonal subspaces and irreversible time evolution.   } }
 \end{center}
 \end{figure}

The subspace $\Theta$ may represent a resonance (or generally an unstable object) decaying into the given particle pair. The unstable object may be, of course, formed also in the decays of heavier unstable objects. In such a case the decay process may be represented always as the transition from $\Theta$ to  $\Delta^+$. 
  
In the case of discrete Hamiltonian spectrum (e.g., harmonic
oscillator) the wave function has similar irreversible t-dependent form.
However, the evolution is periodical as a rule.  The evolution may be again characterized by trajectories
corresponding to different initial conditions. 
The corresponding Hilbert space consists then of infinite number of identical pairs of mutually orthogonal subspaces belonging subsequently to two intervals of phase: $(k.\pi, (k+1).\pi)$ and $((k+1).\pi,(k+2).\pi)$; individual states being characterized by expectation values of phase $\Phi$ lying in principle in the interval from $-\infty$ to $+\infty$ (see also \cite{lok98}).

Such an extended model enables to represent the time evolution described by Schr\"{o}dinger equation in the Hilbert space in full agreement with actual behavior of physical systems. It removes fully the criticism of Pauli; however, it does not give any answer to the criticism of Susskind and Glogover. To this goal some other extension of Hilbert space would be necessary as introduced in Sec. \ref{sec3}. 

As to the irreversibility of time evolution it corresponds to the behavior of real objects in microscopic world similarly as it has been in macroscopic one. There is not practically any greater physical gap between these two regions. 

\section{ EPR experiment and HV theory}
\label{sec5}
In the early years of quantum mechanics only the Copenhagen alternative was taken as valid since the physical community was influenced strongly by the "proof" of von Neumann \cite{vonne} that any other (hidden) parameters were excluded by the standard quantum-mechanical mathematical model and that the given quantum-mechanical model represented complete description of microscopic world. It remained without any attention that already in 1935 Grete Herrmann \cite{gher} showed that the approach of von Neumann was practically a "circle proof".

It was undoubtedly the main reason why also the criticism of Einstein and his collaborators \cite{einst} was refused. Einstein proposed the known EPR Gedankenexperiment and argued in 1935 that the quantum mechanics was not a complete theory to describe fully a physical system. 
Bohr \cite{bohr} opposed strongly; having stated that the Copenhagen model corresponded fully to microscopic reality. And physical community accepted practically his standpoint.

The argument that a hidden variable was contained already in Schr\"{o}dinger equation was published by D. Bohm \cite{bohm} in 1952; however, it was accepted seriously by a very small number of the then physicists. Only the approach of J. Bell \cite{bell} met with greater attention, as also some formulas were presented that seemed to be able to bring the decision between the two discussed (Copenhagen and ensemble) alternatives of the quantum-mechanical model on experimental basis.

However, in broad physical community there has not been any interest to change generally accepted paradigm. Any greater doubts about the standard quantum theory have not been evoked, which was influenced from the very beginning in principle by the mentioned mistaking proof of von Neumann. 

Nevertheless, the approach initiated by Bell influenced decisively the whole further story of the EPR experiment. The original Gedankenexperiment was modified to be possible to perform it. And the corresponding experiments started to be prepared. However, two other mistaking arguments played a decisive role in the then solution of the given problem. 

One argument followed from the statement of Belinfante \cite{belin} that the Copenhagen alternative and the HV theory had to give mutually different predictions, or in other words, that the prediction of HV theory had to differ significantly from the Malus law (approximately measured dependence of light transmission in the case of two polarizers). That was not, however, true as the given statement was based on mistaking interchange of transmission probabilities through a polarizer pair and one polarizer; a more detailed explanation having been given, e.g., in Refs. \cite{lk02,contr}. In fact, the approximate Malus law (as measured) has been fully consistent with the HV theory.

The other mistaking argument was then involved in the already mentioned formula of Bell, which will be explained to a greater detail in the next section. Now the actual story of the EPR experiment will be described. 

 The experimentally feasible EPR experiment consisted in the measurement of coincidence transmission probabilities of two photons with opposite spins and running in opposite directions through two polarizers:
            \[   <---|^{\beta}---o---|^{\alpha}--->  \]
where $\alpha$ and $\beta$ are deviations of individual polarizer axes from a common zero position.
According to Bell \cite{bell} any four transmission probabilities $a_j, b_j \;(j=1,2)$ corresponding to two different orientations of both the polarizers  (4 different combinations) should have fulfilled the following condition
           \[  B = a_1b_1+a_2b_1+a_1b_2-a_2b_2 \leq 2   \]  
if the HV theory had hold (which has not been, however, true as it will be shown in Sec. \ref{sec6}). For the Copenhagen alternative this upper limit should have been higher.
The then situation was described to a greater detail in already mentioned book of
Belinfante \cite{belin}.

The main series of corresponding EPR experiments was being performed
since 1971. It was finished practically in 1982 by the
following conclusion (see Aspect et al. \cite{asp}):

 -  Bell's inequalities have been violated; and the hidden-variable
 alternative seemed, therefore, to be refused;

 -  measured values have been practically in agreement with
 quantum-mechanical predictions (approximate Malus law having been obtained).

At that time the results were interpreted as the victory of Copenhagen
quantum mechanics. However, it was seen soon that practically
nothing from earlier problems was solved. And the discussions
concerning EPR experiments have continued. In fact none of both the results of Aspect's experiment has led to refusal of HV theory as the corresponding conclusion has been based on the mentioned mistakes.

As to the latter result of Aspect et al. obtained in the given EPR experiment (approximate Malus law) we have already mentioned the mistake of Belinfante and referred to its detailed explanation in Ref. \cite{contr}. The approximate Malus law may correspond fully to both the theoretical alternatives and no such an argument against the HV theory exists.

More detailed explanation is needed, of course, as to the first experimental result. Bell started in principle from the HV approach. However, to derive the given inequalities he had to interchange some probabilities, which seemed to be natural at the first sight only. In fact, it represented a strong assumption that corresponded to the passage from the hidden-variable concept to the classical one. The given inequalities were derived, of course, in other ways, too; see, e.g., Ref. \cite{clau}. Similar assumptions were involved, however, in all these approaches (see \cite{lk98}). Bell's limit corresponded to classical physics and not to HV theory, which will be discussed and explained to a greater detail in the next section.

However, at the end of this section it is necessary to mention also the recent delayed-choice experiment \cite {jac} trying to bring a new support for Copenhagen alternative. It follows from our theoretical analysis that no further experiment of EPR type may influence our conclusions concerning the decisive preference of the hidden-variable theory.
 
\section{ Bell's operator and different inequalities }
\label{sec6}
It is evident from the preceding that Bell's combination of
coincidence probabilities may exhibit different limits according
to basic assumptions concerning the individual processes. We will discuss this problem now in the language of the so called
Bell operator obtained by substituting individual probabilities by basic operators representing individual measurement acts (see, e.g., \cite{hill}). According to chosen assumptions  three different limits may be obtained.

The Bell operator $B$ may be represented in the Hilbert space
 \begin{equation}
     {\mathcal H}\;=\; {\mathcal H}_a \otimes {\mathcal H}_b
                         \label{tens}
 \end{equation}
where the subspaces ${\mathcal H}_a$ and ${\mathcal H}_b$
correspond to individual measuring devices (polarizers) in the
coincidence arrangement. It is then possible to introduce the
operator
 \begin{equation}
    B\;=\;a_1b_1+a_1b_2+a_2b_1-a_2b_2
 \end{equation}
where $a_j$ and $b_k$ are now operators acting in subspaces ${\mathcal
H}_a$ and ${\mathcal H}_b$ and corresponding to different
measurements in individual polarizers. It holds for the
expectation values of these operators \cite{hill}
   $$ 0\;\leq\; |\langle a_j\rangle|, \;|\langle b_k\rangle|\; \leq \;1\, .  $$
The expectation values $|\langle B\rangle|$ of the Bell operator
may then possess  different upper limits according to the mutual
commutation relations of the operators $a_j$ and $b_k$.

 If it holds
     $$  [a_j,b_k]\neq 0, \;\;[a_1,a_2]\neq 0, \;\;[b_1,b_2]\neq 0    $$
one can obtain by a rough estimate \cite{tsil}
   $$ \langle BB^+\rangle \leq 16 \;\; \mathrm {or}  \;\;\langle B\rangle \leq 4 \;. $$
However, after more detailed calculation one obtains (see
\cite{revz})
 \begin{equation}
 \langle BB^+\rangle \leq 12, \;\;\;\;|\langle B\rangle| \leq 2{\sqrt 3}\; . \label{one}
 \end{equation}
If
 $$ [a_j,b_k]= 0\;,\;\;\mathrm {and}\;\;\;[a_1,a_2]\neq 0,\;[b_1,b_2]\neq 0\,,     $$
it holds
 \begin{equation}
   \langle BB^+\rangle \leq 8,
     \;\;\;\;|\langle B\rangle| \leq 2{\sqrt 2}  \;.  \label{two}
 \end{equation}
And finally, if all operators $a_j$ and $b_k$ commute mutually one
obtains
 \begin{equation}
   \langle BB^+\rangle \leq 4,
              \;\;\;\;|\langle B\rangle| \leq 2 \; ;  \label{three}
 \end{equation}
the same limit being obtained also if at least the operators
belonging to one of subspaces ${\mathcal H}_a$ or ${\mathcal
H}_b$ commute mutually and with all other operators \cite{revz}.

It has been, therefore, derived that there are in principle three different limiting bounds for Bell's combination of coincidence probability measurements. In the following we will attempt to correlate them to individual physical alternatives:

(i) In contradistinction to hitherto common opinion the last limit
(\ref{three}) corresponds to the conditions of classical physics. 

(ii) The limit (\ref{two}) represents the properties of the HV alternative. There are not, of course, any hidden parameters; all physical parameters are standard ones; only some of them being statistically distributed in corresponding initial states.

(iii) As to the limit (\ref{one}) it represents the case when the
results of both the measuring devices may influence mutually one
another; it would be the case of the orthodox (Copenhagen) quantum mechanics.

It is only the classical limit that has been excluded by experimental EPR data. As to the HV alternative it does not contradict
the experimental results and may be brought to agreement with
experimentally established coincidence polarization data
(obtained, e.g., by Aspect et al. \cite{asp}). It is, of course, also the
Copenhagen quantum mechanics that has not been excluded by EPR experiment results. Here the previously discussed internal discrepancies must be taken into account.

\section{ HV theory and physical reality }
\label{sec7}
It follows from the preceding that there are not any objections against the HV theory as to the description of microscopic world. However, the introduced arguments indicate that this theory should be not only preferred to the standard quantum mechanics but also used in describing all physical reality, as it has been considered recently by A. Legget \cite{legg,legg2}.
He has asked whether the every day world may be described with the help of the same physical model as microscopic objects.
 However, he has not taken into account the significant difference between two alternatives of  quantum-theoretical model: statistical (or HV theory) and Copenhagen. Consequently, he has not been able to get a positive answer. 
His question has been answered, of course, positively in our paper, as the former alternative, i.e. HV theoretical approach, differs from the classical physics only in the existence of discrete energy spectrum in closed physical systems (see also Sec. \ref{sec9}). 

The HV theory may be applied, therefore, practically also to the description of standard macroscopic processes, since in the case of discrete spectrum the differences between individual energy values should remain quite immeasurable. However, quite recent analyses concerning the cavity optomechanical systems seem to provide the way to much stronger support for the HV theory to be denoted as generally valid; see e.g. \cite{thom,bhatt}. The behavior analogous to the excitations of single molecules may be expected for pairs of membranes in vibrational states \cite{bhatt}. However, it is not possible to interpret such states as the phenomenon of entanglement (see Ref. \cite{hart}) as this phenomenon might be bound only to the Copenhagen quantum mechanics that contains internal contradictions. The application of the HV theory might be probably very helpful in the solution of corresponding problems of optomechanics.   

It is also the famous problem of Schr\"{o}dinger cat (representing a macroscopic closed physical system) that may be seen in a quite new light now. There are two basic states ("eigenstates"), i.e., living and dead cat, and any superposition of these states correspond to their statistical combination.

\section{ Light transmission through three polarizers }
\label{sec8}
In the preceding we have discussed the internal discrepancies and unphysical aspects of the Copenhagen quantum mechanics and demonstrated that they might be removed by passing to the HV theory. At the same time the criticism of Einstein may be removed, too. Any objections against HV theory do not exist.

In the case of the EPR experiment with polarized photons both the quantum theories may give practically the same predictions for coincidence transmission, even if the physical interpretations of polarization and transmission mechanisms are quite different. And one may ask whether different predictions may exist for another arrangement of polarizers. Already some time ago we have attempted to analyze the transmission of light through three polarizers:
               \[       o---|---|^{\alpha}---|^{\beta}--->    \]
where individual polarizers have been differently oriented;  $\alpha$  and  $\beta$ denoting axis deviations of the second and third polarizers from the first one. According to the Copenhagen quantum-mechanical model it should hold for corresponding light transmission probability
\begin{equation}
       P(\alpha,\beta)\,=\,\cos^2\!\alpha\;cos^2(\alpha-\beta). \label{Cqm}
\end{equation}
And it was possible to expect measurable deviations in the case of the HV theory.

The corresponding experiment was performed and the results were published in 1993-4; see \cite{krasa1,krasa2}. For a given angle $\alpha$ the angle $\beta$  was always established, so as the total light transmission be minimal. Fundamental deviations from the given quantum-mechanical formula (\ref{Cqm}) have been found, as it may be seen from Figs. 2 and 3.

The mentioned angle pairs (giving the minimum light transmission for given $\alpha$) are shown in Fig. 2. And the corresponding experimental values of transmitted light are represented by dashed line in Fig. 3 (experimental points taken from \cite{krasa2}). The quantum-mechanical predictions for the given angle pairs are then represented by full line; the position of this line being shifted in vertical
direction somewhat arbitrarily as the values of the so called "imperfectness" of given polarizers were not available. In any case the standard quantum-mechanical prediction requires maximums at the positions where the experiment exhibits deep minimums. The difference against the Copenhagen quantum mechanics was not accented in Ref. \cite{krasa2} where the experimental results were published for the first time as we were afraid reasonably that the paper would not have been accepted for publication in such a case. We have only mentioned that similar characteristics may be obtained in the framework of classical theory of Stokes.  

And it is possible to conclude that the given experimental data represent further falsification of Copenhagen quantum mechanics, while good agreement may be evidently obtained in the framework of HV theory as a series of free parameters for the polarization description is available.

\begin{figure}[t!]
\begin{center}
\includegraphics*[scale=.40, angle=-90]{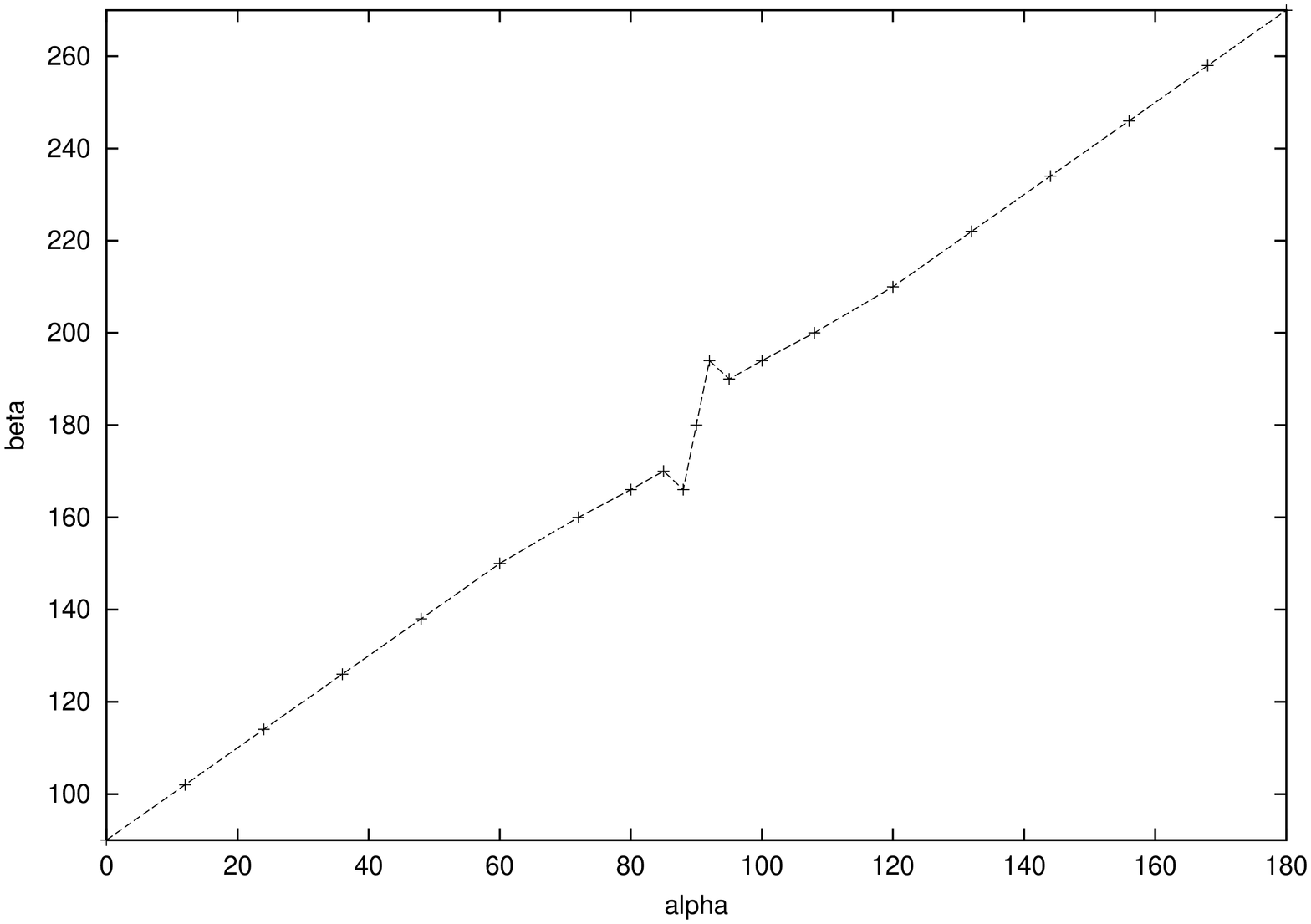}
\caption { \it {  Pairs of angles $\,\alpha$ and $\beta\,$ corresponding to minimal transmission values for chosen $\alpha$;
         pair values used for the measurement shown in Fig. 3.  } }
 \end{center}
\vspace{4mm}
\begin{center}
\includegraphics*[scale=.40, angle=-90]{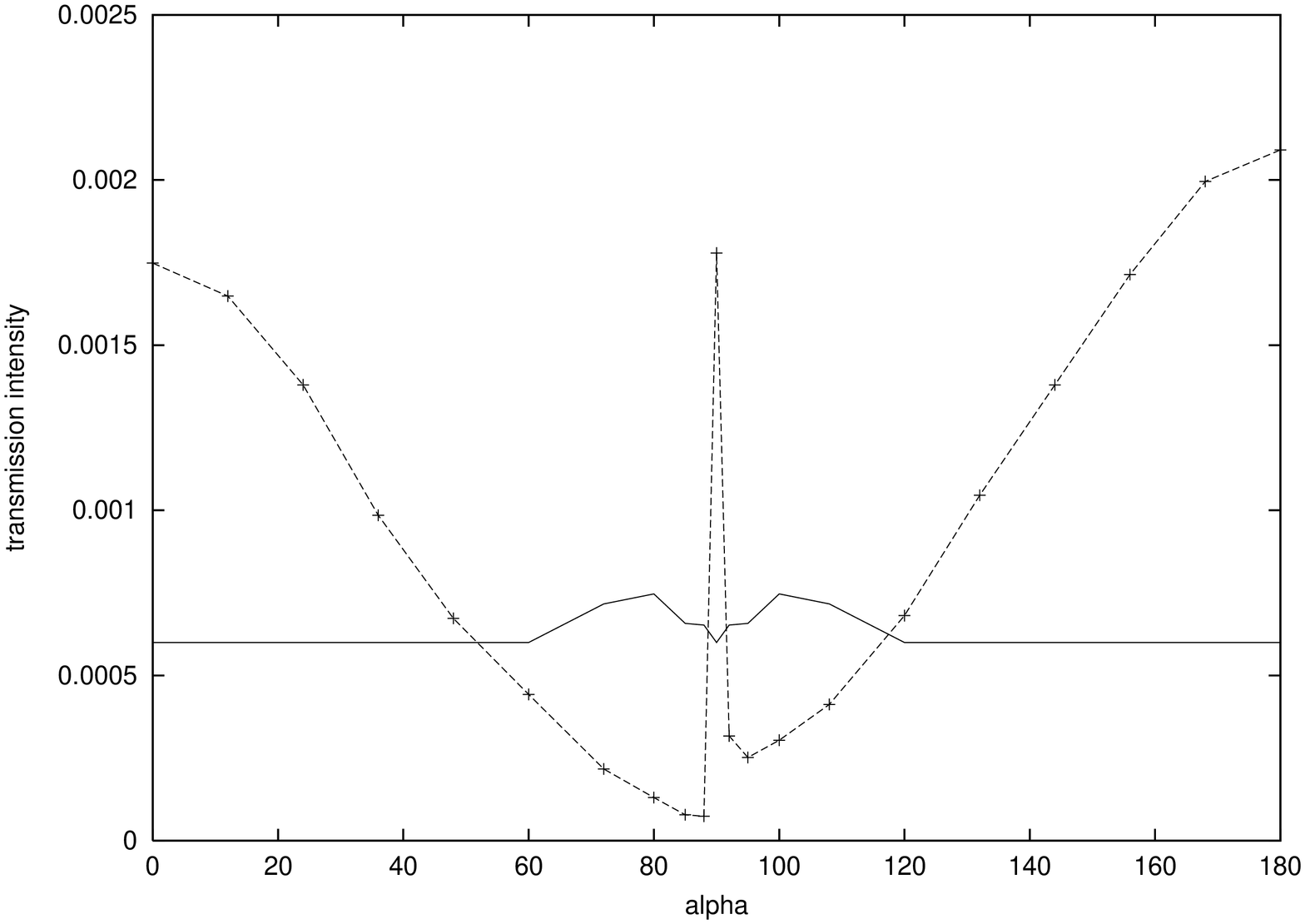}
  \caption { \it { Light transmission through three polarizers (for
     $\,\alpha\,-\,\beta\,$ pairs shown in Fig. 2); experimental data -
     points on dashed line; quantum-mechanical (orthodox) prediction -
     full line.     } }
 \end{center}
 \end{figure}

\section{ Schr\"{o}dinger equation and classical physics }
\label{sec9}
To complete our analysis it is necessary to discuss the actual relation between the Schr\"{o}dinger equation and classical physics, yet. E. Schr\"{o}dinger \cite{schr} was successful with his equation when he showed that for particles exhibiting inertial motion the identical behavior with classical physics was obtained. 
Let us return now, therefore, to the time-dependent Schr\"{o}dinger equation
\begin{equation}
              i\hbar\frac{\partial}{\partial t}\psi(x,t)=H\psi(x,t)   \label{tsch}
\end{equation}
where the complex function $\psi(x,t)$ is expressed as
\begin{equation}
        \psi(x,t)  \;=\; \lambda(x,t)\, e^{\frac{i}{\hbar}\Phi(x,t)}  \label{psi}
\end{equation}
and both the functions $\lambda(x,t)$ and $\Phi(x,t)$ are real. Let us limit to time-independent potential $V(x)$ (see Eq. (\ref{schr})).  Eq. (\ref{tsch}) may be substituted by two equations for two real functions (see D. Bohm \cite{bohm})  
\begin{eqnarray}
 \frac{(\nabla \Phi)^2}{2m}\,+\, V(x)\,+\,V_q(x,t)
                       &=& -\,\partial_t\,\Phi \;,   \label{hamj}   \\
 \triangle\Phi \,+\,2(\nabla\,\Phi)(\nabla\,lg\,\lambda) &=&
       -2m\;\partial_t\, lg\,\lambda   \label{ham2}
\end{eqnarray}
where
\begin{equation}
     V_q(x,t)\,=\,-\frac{\hbar^2}{2m}\frac{\triangle\lambda}{\lambda}
\end{equation}
has been denoted as quantum potential.   

Eq. (\ref{hamj}) resembles Hamilton-Jacobi equation 
\begin{equation}
      \frac{1}{2m}(\nabla S(x,t))^2 + V(x) = -{\partial_t S(x,t)}   \label{haja}
\end{equation}
where $S(x,t)$ has been replaced by $\Phi(x,t)$ and the quantum potential $V_q(x,t)$ has been added.  $S(x,t)$ is Hamilton principal function, from which the momentum values may be derived:
                \[   p(x,t)=\nabla S(x,t).  \]
For inertial motion it holds 
               \[ V_q(x,t)\,=\, V(x)\,=\,0  \]
and                               
              \[  \Phi(x,t)\,=\, S(x,t);  \]
the phase being identical with Hamilton principal function in such a case.              

Let us assume now
                       \[ V(x)\; \neq \; 0  \]
and let us limit to basic solutions corresponding to different values of energy (Hamiltonian eigenvalues) and fulfilling the conditions
        \[  \psi^{(E)}(x,t)\,=\,\psi_E(x)e^{-iEt},\;\;\; H\psi_E(x)\,=\,E\psi_E(x).  \]
In such a case $V_q(x)\neq 0$ is independent of $t$ and $\Phi(x,t)$ and $S(x,t)$ are mutually different; $V_q(x)$ representing the numerical measure of such difference.            
 There is not, however, any difference between the physical results of Schr\"{o}dinger equation and classical physics; see also \cite{adv}. All basic solutions of Schr\"{o}dinger equation are fully equivalent to classical solutions corresponding to the same energy. However, it does not hold in opposite direction. For some solutions of Hamilton equation the corresponding counterparts in the Schr\"{o}dinger equation (or in the HV theory) do not seem to exist in the case of discrete spectrum.

In the past when the existence of quantum potential was assumed to represent decisive physical difference between Schr\"{o}dinger equation and classical physics there were done some attempts to interpret it as the consequence of Brown motions of individual microscopic objects. Our result is, however, in full agreement with results of Ioanidou \cite{ioan} and Hoyer \cite{hoyer} who have shown that Schr\"{o}dinger equation may be derived if classical physics is combined with a kind of statistical distribution. 
   
The advantage of the Schr\"{o}dinger equation consists then in obtaining a complete statistical result in one solution if a statistical distribution of initial basic states is given. Consequently, the Schr\"{o}dinger equation is very suitable if some initial parameters (e.g., the impact parameter in collision processes and, consequently, also interaction energy values) are not exactly defined and only their statistical distributions may be established, as it occurs in collision measurement approaches.  

\section{ HV theory and Hilbert space }
\label{sec10}
It follows from the preceding that individual solutions of Schr\"{o}dinger equation may be truly represented in the Hilbert space  that is extended (i.e., at least doubled) in comparison to the third assumption introduced in Sec. \ref{sec2}; and when the fourth assumption is refused, too. The evolution of a physical system is then characterized by a trajectory that represents irreversible behavior. Exact solutions may be derived, of course, practically in the case of a system consisting of two stable particles. However, the representation of time evolution in a corresponding Hilbert space may be very helpful also in the case of more complex physical systems.

Our considerations will be based on the analysis of a two-particle system as it has been described in Sec. 4. Such a scheme may represent a basic structure for describing the evolution of any more complex physical system even if it involves objects that are not stable. A resonance may be formed in the collision of two simpler particles and also the decay of an unstable object (resonance) may be interpreted at least in the first approximation as the transition to two-particle system even if any of the arising particles may be unstable. In such a case the Hilbert space must be, of course, more complex consisting not only of orthogonal sums but also of tensor products of simpler subspaces.

It is clear that all subspaces in one orthogonal sum ($\Delta^+,\Delta^-,\Theta$) must correspond to the same quantum numbers that must be conserved during the whole evolution. One two-particle system at a given time may be then represented in $\Delta^\pm$ by one vector. However, even the stable objects exhibit usually some internal structures and internal evolutions that might influence the transition probabilities to other subspaces in collision processes. It may be taken into account by substituting, e.g., $\Delta ^-$ by the tensor product of Hilbert subspaces $\Delta ^-\otimes{\mathcal P}_1\otimes{\mathcal P}_2$ where the two latter subspaces represent main properties of individual objects. 

It would be, of course, difficult to allocate a more general Schr\"{o}dinger equation to such a physical system. In fact, it is hardly possible to describe the detailed evolution of corresponding collision processes when the detailed internal structures and evolutions of individual particles are not known. It is possible to characterize the influence of some characteristics only by establishing some probability distributions in transmission processes. And in such a case the representations of ${\mathcal P}^\pm$ may consist in characterizing them by some basis vectors corresponding to different classes of particle properties that may change during the time evolution.
And it is evident that even rather complicated processes might be represented correspondingly (at least at not very high energies). 

The same holds before all also for the subspace $\Theta$ representing a resonance (or an unstable object); it may be suitable to represent it, e.g., by vector basis corresponding to individual decay channels (see \cite{lk69}), where the frequencies of individual basic states may be derived from experimental transmission data.   

\section{ Conclusion }
\label{sec11}
The physics of microscopic world in the 20th century has been represented by the Copenhagen quantum mechanics with its logical paradoxes and contradictions. The given theory was taken as valid during the whole century even if Pauli called the attention to one important contradiction already in 1933: the Hamiltonian had to possess the continuous spectrum in the whole interval of all real numbers $(-\infty,+\infty)$ while the actual energy value must be practically positive. Other critical arguments including also the disagreement with experimental data (light transmission through three polarizers) have been then summarized in the preceding.

It has been shown that all known critical points can be removed when one passes from the Copenhagen quantum mechanics to the HV theory that is based practically on mere Schr\"{o}dinger equation, while earlier additional assumptions (see Sec. \ref{sec2}) are abandoned and  the Hilbert space is adapted to actual time-dependent Schr\"{o}dinger solutions. E.g., for the system of two particles it has been necessary to extend the Hilbert space so as to consist at least of two mutually orthogonal subspaces (see Sec. \ref{sec4}).

The given passage to the HV theory has solved practically all known problems; the mistaking statements having been repaired. It has been also the EPR experiment that should be (and may be) interpreted as it was required by Einstein. 
It has been shown also that the Schr\"{o}dinger equation or the HV theory is practically equivalent to classical physics with the only exception, concerning the existence of discrete bound states. 

The HV theory may be then applied in principle to the description of microscopic as well as macroscopic objects. There is practically no gap between these two physical regions. It is possible also to say that the question put by A. Legget (see Ref. \cite{legg2}) has been answered in positive way only if one has passed from Copenhagen quantum mechanics to HV theory (see also Sec. \ref{sec7}).

{\footnotesize

\end{document}